\documentstyle[12pt]{article}
\def\beq{\begin{eqnarray}}    
\def\eeq{\end{eqnarray}}      

\def\Tr{\,\mbox{Tr}\,}                  

\def\de{\delta}

\def\na{\nabla}

\def\si{\sigma}

\def\ph{\varphi}

\def\Ga{\Gamma}
\def\De{\Delta}

\begin{document}
\begin{center}

{\Large\sc
Vacuum effective action and inflation}
\vskip 4mm

{\bf  J.C. Fabris $^a$}
 \footnote{Electronic address: fabris@cce.ufes.br},$\,$
{\bf A.M. Pelinson $^b$}
 \footnote{Electronic address: ana@fisica.ufjf.br},$\,$
{\bf I.L. Shapiro $^{b}$}
 \footnote{On leave from Tomsk Pedagogical University,
Tomsk, Russia. E-mail: shapiro@fisica.ufjf.br},$\,$
\vskip 3mm

a. {\small Departamento de F\'{\i}sica -- CCE,
Universidade Federal de Espir\'{\i}to Santo, ES, Brazil}

b. {\small Departamento de F\'{\i}sica -- ICE,
Universidade Federal de Juiz de Fora, MG, Brazil}
\vskip 2mm
\end{center}

\vskip 6mm

\noindent
{\large\it Abstract}
\footnote{Contribution to the Proceedings of the X Jorge 
Andre Swieca school in Particles and Fields. To be published 
in World Scientific.}.
$\,\,${\sl
We consider vacuum quantum effects in the Early Universe,
which may lead to inflation. The inflation is a
direct consequence of the supposition that, at high energies,
all the particles can be described by the weakly interacting,
massless, conformally invariant fields. We discuss, from the
effective field theory point of view, the stability of inflation,
transition to the FRW solution, and also possibility to
study metric and density perturbations.}
\vskip 6mm

The aim of the present article is to discuss the importance
of the quantum
effects of vacuum for the evolution of the Early Universe.
In particular, we shall present a strong arguments
that the effective action
of vacuum induced by conformal anomaly produces inflation in
a very natural way. This problem was previously discussed
in \cite{fhh,mamo} and studied in details in
\cite{star1,star2} (see also \cite{anju}),
but we hope to shed some more light on it and also draw
some prospects and frameworks for the future studies.

Let us start from the
description of possible physical situation
in the Early Universe, in which the trace anomaly plays the
leading role. The main supposition is that there
is a desert in the particle spectrum between the Planck mass
$M_P \approx 10^{19} GeV$ and the heaviest massive
particle. This assumption has serious backgrounds, because
the existing particle theories indicate that the unification
point lies as $M_X \approx 10^{14} - 10^{16} GeV$. For greater
simplicity one can suppose that the
$U(1)\times SU(2)_L \times SU(3)$ Standard Model of the
Particle Physics is valid till $M_P$ scale, that makes
a desired desert very large. Then, all the matter fields
below the $M_P$ scale can be described by the effective
local quantum field theory. Moreover, if the desert is
long enough, at high energies all the particles
can be safely treated as massless. Then we
can apply the standard relation $\,p=1/3\,\rho\,$
between pressure and energy density for the massless particles.
If we turn to the field description, this relation indicates
that all the fields possess not only global, but
also local conformal invariance \cite{anju}. On the other
hand, for very large kinetic energy, the potential energy of
the interactions between quantum fields is disregardable, except
the vacuum effects. Thus, the most
natural model for the Early Universe is the curved
manifold filled by the free, massless, conformal
invariant fields. Suppose, the underlying matter
theory has $N_0$ scalars, $N_{1/2}$ Dirac spinors
(Weyl spinors may be treated in a similar way),
and $N_1$ vectors. The quantum effects of these
fields on curved background will be the object of our
present study.

The classical solution for the pure radiation case
is usual $\,a \sim t^{1/2}$. However, in the high energy
region below $M_P$ the vacuum quantum effects of
matter fields must be taken into
account, and the geometry of the background may change.
For the situation described above, the
leading quantum effect is the one related to the
renormalization of the vacuum action \cite{birdav,book}.
This action includes, for the conformal case, three
necessary structures (conformal invariant
and surface terms):
\beq
S_{vac} = \int d^4 x\sqrt{- g}\,
\left\{l_1C^2 + l_2E + l_3{\Box}R\,\right\}\,,
\label{vacuum}
\eeq
where $l_{1,2,3}$ are some parameters, $C^2$ is the
square of the Weyl tensor and $E$ is the
integrand of the Gauss-Bonnet topological (in $d=4$) invariant.
One has to notice that the introduction of the
non-conformal terms like $\,\int\sqrt{-g}R\,$ or
$\,\int\sqrt{-g}R^2\,$ is possible but not necessary for the
renormalization of the free conformal invariant theories.
In other words, non-conformal terms do not renormalize.

The renormalization of the action (\ref{vacuum})
leads to conformal anomaly \cite{duff} and to the
anomaly-induced effective action \cite{rei,frts}
(see also \cite{book}).
This action can be written in the form:
$$
{\bar \Ga} = S_c[{\bar g}_{\mu\nu}] +
\int d^4 x\sqrt{-{\bar g}}\,\{
w\si {\bar C}^2 + b\si ({\bar E}-\frac23 {\bar {\Box}}
{\bar R}) + 2b\si{\bar \De}\si -
$$
\beq
- \frac{3c+2b}{36}\,
[{\bar R} - 6({\bar \na}\si)^2 -
6{\bar \Box} \si]^2)\}.
\label{quantum}
\eeq
Here ${g}_{\mu\nu} = a^2(\eta)\cdot {\bar g}_{\mu\nu}\,,\,\,\,\,
\sigma(\eta) = \ln a(\eta)$, and
the fiducial metric ${\bar g}_{\mu\nu}$ has fixed determinant.
$\,\De\,$ is conformal invariant self-adjoint operator
$$
\De = {\Box}^2 + 2R^{\mu\nu}\na_\mu\na_\nu
-\frac23\,R{\Box}+\frac13\,(\na^\mu R)\na_\mu\,,
$$
coefficients $w,b,c$ are
\beq
w=\frac{N_0 + 6N_{1/2} + 12N_1}{120\cdot (4\pi)^2 }\,\,,
\,\,\,\,\,\,\,
b= -\,\frac{N_0 +
11N_{1/2} + 62N_1}{360\cdot (4\pi)^2 }\,\,,
\,\,\,\,\,\,\,
c=\frac{N_0 + 6N_{1/2} - 18N_1}{180\cdot (4\pi)^2}\,,
\label{abc}
\eeq
and $S_c[g_{\mu\nu}]$ is some unknown conformal-invariant
functional. In general, exact calculation of this functional is
impossible. As far as we are interested in the conformally flat
cosmological solutions, this functional has no importance.
However it becomes relevant for we intend to study
some non-isotropy, and in particular explore metric
perturbations. The situation with the perturbations is
quite similar to the one for the
black-hole physics. In the last case one can apply
the Reigert solution (\ref{quantum}) and achieve "correct"
result for the Hawking radiation, but only for the
particular choice of $\,\,S_c[g_{\mu\nu}]\,\,$ \cite{balbi}.

It is important to notice that the
solution (\ref{quantum}) can be rewritten in the covariant
but non-local form \cite{rei}. In turn, this can be transformed
again into local expression depending on two
auxiliary scalar fields $\ph$ and $\psi\,\,$ \cite{rei,balbi}.
$$
\Gamma = S_c[g_{\mu\nu}] -
\frac{c-\frac23\,b}{12\,(4\pi)^2}\,\int d^4 x \sqrt{-g (x)}\,R^2(x) +
 \int d^4 x \sqrt{-g (x)}\,\left\{\,
\frac12 \,\ph\De\ph - \frac12 \,\psi\De\psi+
\right.
$$
\beq
\left.
+ \ph\,\left[\,\frac{\sqrt{-b}}{8\pi}\,(E -\frac23\,{\Box}R)\,
- \frac{w}{4\pi\sqrt{-b}}\,C^2\,\right]
+ \frac{w}{4\pi \sqrt{-b}}\,\psi\,C^2\,\right\}\,.
\label{finaction}
\eeq
Indeed this local covariant
formulation is more suitable for the study of metric
perturbations. However the isotropic solution
can be achieved on the basis of the action (\ref{quantum}).
For the Early Universe the curvature (positive or negative)
of the space section of the space-time has no importance,
and thus one can restrict consideration by the
conformally flat case.

Let us derive the equation of motion for the $\sigma(\eta)$,
put ${\bar g}_{\mu\nu} = \eta_{\mu\nu}$ and pass to the
new variables: physical time $d\eta = a(\eta)\cdot dt\,$
and $H(t)= {\dot a}(t)/a(t)$. The equation for $H(t)$
has the form \cite{anju}
\beq
{\stackrel{...} {H}}
+ 7 {\stackrel{..} {H}}H
+ 4\,\left(3 - \frac{b}{c}\right)\,
{\stackrel{.} {H}}H^2 + 4\,{{\stackrel{.} {H}}}^2
- 4\,\frac{b}{c}\,H^4 - \frac{2M^2_{Pl}}{c}\,
\left(\,H^2 + {\stackrel{.} {H}}\,\right)  = 0\,,
\label{logs}
\eeq
The special inflationary solution corresponds to $\,H = const$:
\beq
H = \pm \frac{M_P}{\sqrt{-b}}\,,\,\,\,\,\,\,\,\,\,\,\,\,\,\,
a(t) = a_0\cdot \exp {Ht}\,,
\label{inflation}
\eeq
positive sign corresponds to inflation. The solution (\ref{inflation})
was first discovered and studied in \cite{mamo,star1,star2}
\footnote{In \cite{star1} two
other similar solutions for the FRW metric with $k=\pm 1$
were found.}.
Here we shall discuss the behaviour of the $a(t)$
on the basis of the effective approach to quantum theory.
The inflationary solution for the local covariant version
of the induced action (\ref{finaction}) can
be achieved, and the behaviour of the conformal factor is
the same as in (\ref{inflation}). At the same time
the inflationary solution of (\ref{finaction})
contains the arbitrariness related to the boundary conditions
for the auxiliary fields $\ph,\psi$. We remark that in the
black-hole case this arbitrariness allows one to classify the
vacuum states \cite{balbi}, and therefore it is natural to expect
that it can be successfully used in the study of the metric
perturbations around the solution (\ref{inflation}).

The detailed analysis shows \cite{star1,anju} that the
special solution is stable with respect to the variations
of $a(t)$, if the
parameters of the underlying quantum theory satisfy the
condition $\frac{b}{c} < 0$, that leads, according to (\ref{abc}),
to the relation\footnote{This constraint is not satisfied
for the Minimal Standard Model (MSM), but it is may be easily
achieved in some generalizations including the supersymmetric
MSM. The constraint imposed in \cite{anselmi} also provides
stability of inflation.}
\beq
N_1\, <\, \frac13\,N_{1/2}\, + \,\frac{1}{18}\, N_0\,.
\label{const}
\eeq
With this constraint
satisfied, the theory goes to inflation independent on the
initial conditions, and the inflation is eternal -- unless
the masses of the particles become seen and change the
structure of effective action. We remark that
the coefficient $c$ in (\ref{abc}) can be modified by adding finite
$\int\sqrt{-g}R^2$-term to the vacuum action. This coefficient
does not renormalize in the free theory and has very week
running in the interaction theory \cite{brv}, but it can provide
the stability of inflation for any particle content.
Below we do not consider this term.

Let us remind that the basis for the above solution is an
effective action induced by conformal anomaly -- that is the
quantum effect of the massless conformal invariant fields.
In order to
achieve better understanding of the approximation, let us
suppose that the fields possess masses, and call typical mass
$m$. For simplicity, one can imagine that we have just a
massive scalar with $\xi=1/6$, but the consideration is the
same for the spinor case. One can derive the anomaly through
the Jacobian of the conformal change of variables in the
effective action of vacuum.
\beq
{i\Ga_{vac}} = {iS_{vac}} \,+ \,\ln\,\int\, {\cal D}\ph\,
\exp \{i\,S_{matter}[\ph, g_{\mu\nu}]\}\,.
\label{efac}
\eeq
Then, using the relation
$\,\,\left[\,-2g_{\mu\nu}\,\frac{\de}{\de g_{\mu\nu}} +
\ph\,\frac{\de}{\de \ph} \right]\,S_{matter} \sim m^2\,\,$
we find, for the conformally flat metric, that the
effective action of vacuum is given, in the case, by the sum
\beq
\Ga_{vac}\, = \,{\bar \Ga}\,+\,
\frac{i}{2}\,\Tr\ln\left\{-\Box + m^2\cdot e^{2\si (t)}
\right\}_{g_{\mu\nu}=\eta_{\mu\nu}}\,,
\label{massive}
\eeq
where the last contribution is proportional to the massive
parameter $m^2$. Also, in the matter action, only the massive
part does not decouple from the $a(t)$, and therefore
all "manifestations" of the particle mass is proportional to
the $m^2$. Indeed, the last term in (\ref{massive})
is divergent and leads to the renormalization of the Einstein-Hilbert
and cosmological terms, which should be, in the case, included
to the $S_{vac}$. We remark, that the renormalization of the
cosmological constant leads to its running and therefore to the
appearance of the finite cosmological constant due to the
mechanism recently described in \cite{cosm}. However, this
finite cosmological constant is also proportional to $m^2$.
In the equation for $a(t)$ one can pass to the Feynman graf
description \cite{duffetal}, and then, in the momentum space,
the time derivatives will be substituted by the energy $\mu$.
Then, in the massive case, the equations which govern
the cosmological evolution, have the symbolic form:
\beq
{\cal O}_{ind}(\mu^4) + {\cal O}_{GR}(\mu^2\cdot M_P^2) +
{\cal O}_{massive}(\mu^2\cdot m^2) =  0\,.
\label{eq}
\eeq
At this point one can apply the effective approach to consider
the problem of transition to the power-like FRW Universe.
Consider first
the very early Universe, with $\mu\approx M_{P}$. One can,
according to our "desert hypothesis", suppose that the heaviest
particle has a mass about $100 GeV$. Then the first two terms
in (\ref{eq}) are of the same order, while the last one is
suppresses by the factor of $10^{-34}$. This is the
inflation region, and in case of
(\ref{const}) satisfied we meet universal
behaviour (\ref{inflation}).

Now, let us consider another end of the mass scale, when
$\mu \approx 10^{-12} GeV$ (modern cosmic scale).
Then the first (anomaly induced) term is (\ref{eq})´
is suppressed by the factor of $10^{-28}$ compared to the
last "massive" term, and by the factor of $10^{-62}$
compared to the Einstein part. Indeed, the Universe
does not "see" quantum effects, and evolves by the
usual power low. This simple consideration shows that there
cannot be any unique equation for the Universe, that
could govern its evolution from the start to the end.
Instead, one has to divide the evolution for some
periods which have essentially distinct dynamics.

Indeed, the most complicated is the intermediate period
with $\mu$ comparable to the masses of the matter particles.
However, in this region the
derivation of quantum correction to the effective action
meets serious difficulties. First of all, exact calculation
of the non-local part of the
effective action of the massive fields is unsolved
problem, even within the perturbation theory. Besides, the
QCD vacuum effects are essentially non-perturbative, and
in particular may produce the topological solutions that could
also contribute to the dynamics. Indeed, one can consider some
simplified models (it is natural to regard \cite{fhh}
as the first attempt of this kind) to investigate the
intermediate stage, and thus describe a smooth transition
from inflation to the modern Universe.

The model of inflation described above (it is usually
referred to as the Starobinsky model),
has serious advantages as compared to the
usual models based on the inflaton \cite{review2}.
The main advantage is, of course,
the naturalness of inflation, which appears as unavoidable
consequence of the desert on the mass scale and the
constraint on the particle spectrum (\ref{const}).
On the other hand, in order to compete with the
inflaton models, the theory must pass the series of tests, and in
particular provide the correct density and metric perturbation
spectrum. On the other hand, as it was already mentioned above,
when we study the perturbation, the conformal invariant
functional $S_c[{\bar g}_{\mu\nu}]$ in (\ref{quantum})
becomes relevant, and its proper choice is expected
to provide the desirable result \cite{anj-pro}.

\noindent
{\bf Acknowledgments.} Authors are grateful to CNPq for
the scholarship (A.M.P.) and grants (J.C.F and I.L.Sh.).



\begin {thebibliography}{99}

\bibitem{fhh} M.V. Fischetti, J.B. Hartle and B.L. Hu,
              Phys.Rev. {\bf D20} (1979) 1757.

\bibitem{mamo} S.G. Mamaev and V.M. Mostepanenko, Sov.Phys. -
               JETP {\bf 51} (1980) 9.

\bibitem{star1}
A.A. Starobinski, Phys.Lett. {\bf 91B} (1980) 99.

\bibitem{star2}
A.A. Starobinski, JETP Lett. {\bf 34} (1981) 460; Proceedings of the
second seminar "Quantum Gravity". pg. 58-72. (Moscow, 1981/1982).

\bibitem{anju} J.C. Fabris, A.M. Pelinson, I.L. Shapiro,
Anomaly-induced effective action for gravity and inflation.
gr-qc/9810032.

\bibitem{birdav} N.D. Birell and P.C.W. Davies, {\sl Quantum fields
in curved space} (Cambridge Univ. Press, Cambridge, 1982).

\bibitem{book} I.L. Buchbinder, S.D. Odintsov and I.L. Shapiro,
Effective Action in Quantum Gravity. - IOP Publishing,
(Bristol, 1992).

\bibitem{duff}
S. Deser, M.J. Duff
and C. Isham, {Nucl. Phys.}{\bf 111B} (1976) 45;

M.J. Duff, Nucl. Phys. {\bf 125B} 334 (1977).

\bibitem{rei} R.J. Reigert, Phys.Lett. {\bf 134B} (1980) 56.

\bibitem{frts} 
E.S. Fradkin and A.A. Tseytlin, Phys.Lett. {\bf 134B} (1980) 187.

\bibitem{balbi} R. Balbinot, A. Fabbri and I.L. Shapiro,
Phys.Rev.Lett. {\bf 83} (1999) 1494; $\,\,$  
Nucl.Phys. {\bf B559} (1999) 301.

\bibitem{anselmi} D. Anselmi, {\sl Ann. Phys. (NY} {\bf 276} 
(1999) 361.

\bibitem{brv} G. Cognola and I.L. Shapiro,
Class.Quant.Grav. {\bf 15} (1998) 3411.

\bibitem{cosm} I. L. Shapiro and J. Sol\`{a},
On the scaling behavior of the cosmological
constant and the possible
existence of new forces and new light degrees of freedom.
 hep-ph/9910462.

\bibitem{duffetal}
M.J. Duff, Phys.Rev. {\bf 7D} (1973) 2317; $\,\,${\bf 9D} (1974) 183;

D.G. Boulware and S. Deser, Ann.Phys. {\bf 89} (1975) 193.

\bibitem{review2} E.Kolb and M.Turner, {\sl The very early Universe}
                  (Addison-Wesley, New York, 1994).

\bibitem{anj-pro}J.C. Fabris, A.M. Pelinson, I.L. Shapiro,
Work in progress.

\end{thebibliography}
\end{document}